\documentclass{ws-procs975x65}

\usepackage{amssymb}
\usepackage{amsmath}
\usepackage{amsfonts}
\usepackage{setspace}
\usepackage{color}
\usepackage{subfigure}
\usepackage{float}
\usepackage[utf8]{inputenc}

\numberwithin{equation}{section}

\DeclareMathOperator{\Tr}{Tr}

\def\beq{\begin{equation}}
\def\eeq{\end{equation}}

\begin{document}

\title{Unruh-DeWitt Fermion Detector on a (1+1)-Dimensional Cylindrical Spacetime: Arbitrary Worldlines and Inequivalent Spin Structures}
\author{Jorma Louko$^*$ and Vladimir Toussaint$^\dagger$}

\address{School of Mathematical Sciences, University of Nottingham,\\
Nottingham NG7 2RD, UK\\
$^*$E-mail: jorma.louko@nottingham.ac.uk\\
$^\dagger$E-mail: pmxvt2@nottingham.ac.uk
}

\begin{abstract}
We examine an Unruh-DeWitt particle detector which couples linearly
to the scalar density of a massless Dirac field on the static cylindrical quotient of
the (1+1)-dimensional Minkowski spacetime, allowing the 
detector's motion to remain arbitrary and working to leading order in
perturbation theory.  We show that the detector's response distinguishes the periodic and antiperiodic spin structures, and the zero mode that is present
for periodic spinors contributes to the response by a state-dependent
but well defined and controllable amount. 
\end{abstract}

\bodymatter


\section{Introduction}

Model particle detectors are important mathematical tools used to
explore the properties of quantum fields in flat and curved
spacetimes. These detectors, known in the literature as Unruh-DeWitt
(UDW) detectors, originate in the work of Unruh and DeWitt to analyse
the response of a uniformly accelerated atom in Minkowski spacetime
Refs.~\refcite{Unruh:1976db,Dewitt}. They have been widely used to analyse
effects due to observer motion and spacetime kinematics, including the
Hawking radiation emitted by black holes (see Refs.~\refcite{Takagi:1986kn,birrell-davies,Louko:2007mu,louko-satz-spatial,Barbado:2012fy,Juarez-Aubry:2014jba,Martin-Martinez:2014qda} for a
small selection of references).

In this paper, we analyse an UDW detector which couples linearly
to the scalar density of a massless Dirac field on the static cylindrical quotient of the $(1+1)$-dimensional Minkowski spacetime, allowing the detector's motion to remain arbitrary, to leading order in perturbation theory. We compute explicit expression for the response functions for the inequivalent spin structures  which exhibit no infrared divergence for arbitrary detector's motion along the cylinder, assuming smooth switching.

The main theoretical reasons for considering this model spacetime is firstly  because on the cylinder we have two inequivalent spin structures. We want to find out whether our detector distinguish between the two spin structures. Secondly, the untwisted spin structure contains a mode with vanishing frequency known as a zero mode which does not have a Fock vacuum. We resolve this ambiguity by treating the zero mode as a state with two Fermi degrees of freedom. We then show that the zero mode contributes to the periodic spin structure response by a state-dependent but well defined and controllable amount.  

Throughout we use units in which $\hbar =c =1$, and we work in two-dimensional Minkowski spacetime with signature $(+ -),$ where the minus direction is spacelike. We will replace  $(t,x^1)\equiv x$ whenever we find it more convenient.

\section{Background}
In this section we describe an UDW detector coupled linearly to the
scalar density of the Dirac field of the form $\overline{\psi}\psi$, where $\psi$ satisfies the Dirac field equation, and where the Dirac adjoint field is defined as $\overline{\psi} \equiv \psi^\dagger \gamma^0$.
This model is that of an idealized atom with a set of energy eigenstates $\{|E_0 \rangle , |E_1 \rangle\}$ and which is constrained 
to move along some prescribed classical trajectory:  $t=t(\tau),\ x^1=x^1(\tau)$, where $\tau$ is the detector's proper time. 
The field is weakly coupled to the detector via the interaction Hamiltonian
\begin{equation}
H_{int}= c\chi(\tau)m(\tau)\overline{\psi}\big(x(\tau)\big)\psi\big(x(\tau)\big)
\end{equation}
where $c$ is a coupling constant, $\chi$ is a smooth switching function of finite duration, and $m$ is the detector's 
monopole moment operator. 
Working to leading order in perturbation theory, the detector's transition probability is proportional to the response function, given by    
\begin{align}\label{TransFunct}
\mathcal{F}(\Omega) = \int_{-\infty}^{\infty} \mathrm{d}\tau \int_{-\infty}^{\infty} 
\mathrm{d}\tau^\prime \chi(\tau^\prime) \chi(\tau) e^{-i\Omega(\tau-\tau^\prime)} W^{(2)}(x(\tau), x(\tau^\prime)),
\end{align}
where the transition energy, $\Omega= E_1-E_0$, is such that $\Omega  > 0$ corresponds to excitation while
 $\Omega < 0$ corresponds to de-excitation. The two-point correlation function, assuming that the initial field state is prepared in a (1+1) vacuum state with respect to Minkowski time translation denoted by $ \big |0 \big>$, is
\begin{align}\label{Two_PointFunct}
 W^{(2)}\big(x(\tau), x(\tau^\prime)\big) =  \big \langle 0 \big |\overline{\psi}\big(x(\tau)\big)\psi\big(x(\tau)\big)
\overline{\psi}\big(x(\tau^\prime)\big)\psi\big(x(\tau^\prime)\big)\big| 0 \big \rangle.
\end{align}
Knowledge of this two-point correlation function is required in order to obtain an explicit expression for the detector response function.

\subsection{Discrete Dirac Field Modes for Twisted and Untwisted Spinors}
Consider the cylindrical quotient space of the Minkowski spacetime, with line element
\begin{equation}
 ds^2 =  dt^2 - (dx^1)^2,
 \end{equation}
where $x^1$ is periodic with period $L > 0$. In this spacetime, the complete mode solutions 
to the Dirac field become
\begin{subequations}\label{fieldsmodes}
\begin{eqnarray}
 u_k(t,x^1) &= &(2L\omega)^{-1/2}u_k e^{-i(\omega t - kx^1)},\\
 v_k(t,x^1) &= &(2L\omega)^{-1/2}v_k e^{i(\omega t - kx^1)},
\end{eqnarray}
\end{subequations}
where $\omega_k = \sqrt{m^2+k^2}$, assuming $m > 0$.
Imposing periodic boundary condition, $\psi(t,x^1) = \psi(t,x^1+L)$, we find
\begin{equation}
k_n = \frac{2\pi}{L}n,  \ \ \ n = 0, \pm1, \pm 2, \pm 3...
\end{equation}
Alternatively, we may impose anti-periodic boundary condition,  $\psi(t,x^1) = -\psi(t,x^1+L)$, for which we find
\begin{equation}
k_n = \frac{2\pi}{L}\left(n+\frac{1}{2}\right),  \ \ \ n = 0, \pm1, \pm 2, \pm 3...,
\end{equation}
which is known as the twisted spinor field. 

The Dirac field can now be expanded in terms of the positive and negative frequency modes as
\begin{equation}
\psi(x) = \sum_k\left[ b_ku_k(x) +d_k^{\dagger}v_k(x)      \right] \equiv \psi^+(x) + \psi^-(x) .
\end{equation}
Similarly, the Dirac adjoint field is given by 
\begin{equation}
\overline{\psi}(x)\equiv \psi^\dagger \gamma^0 = \sum_k\left[ b_k^\dagger \overline{u}_k(x) +d_k \overline{v}_k(x)      \right] \equiv \overline{\psi}^-(x)+ \overline{\psi}^+(x)  .
\end{equation}
The field is normalised in such a way that 
\begin{subequations}\label{AntiCommRel}
\begin{eqnarray}
\label{eq.1}\Big\{\psi_a(0,x),\psi_b(0,y) \Big\}&=  &\Big\{\psi_a^{\dagger}(0,x),
\psi_b^{\dagger}(0,y) \Big\} = 0, \\
\label{eq.2}\Big\{\psi_a(0,x),\psi_b(0,y) \Big\} &= &\delta_{ab}\delta(x-y).
\end{eqnarray}
\end{subequations}
 Analogous to the continuous modes,
 we find that for the discrete modes the following relations still hold 
\begin{equation}
\Big\{\overline{\psi}_a^{+}(x), \overline{\psi}_b^{-}(y)   \Big\} = 
\Big\{ \psi_a^{-}(x), \overline{\psi}_b^{-}(y)    \Big\} = 0,
\end{equation}
and
\begin{align}\label{Dis_Wigh_Funct}
\Big\{\psi_a^{+}(x), \overline{\psi}_b^{-}(y) \Big\}&=\bigg[ \big(i\gamma^{\mu}\partial_{\mu}
\big)_x + m   \bigg]_{ab} i\Delta^+(x,y) \nonumber \\
&\equiv i\bigg[S^+(x,y) \bigg]_{ab}  ,
\end{align}
where
\begin{equation}
i\Delta^+(x,y) \equiv \sum_{n=-\infty}^{\infty}\frac{1}{2\omega_nL}e^{-i K_n \cdot(x-y)}.
\end{equation}
And similarly we find,
\begin{align}
\Big\{\psi_a^{-}(x), \overline{\psi}_b^{+}(y) \Big\}&= \bigg[ \big((i\gamma^{\mu}
\partial_{\mu}\big)_x + m   \bigg]_{ab} i\Delta^-(x,y) \nonumber \\
&\equiv i\bigg[S^-(x,y) \bigg]_{ab} ,
\end{align}
where
\begin{equation}
i\Delta^-(x,y) \equiv -\sum_{n=-\infty}^{\infty}\frac{1}{2\omega_nL}e^{i K_n \cdot(x-y)},
\end{equation}
and where
\begin{align}
K_n \cdot(x-y)= \eta_{\mu\nu}K_n^\mu(x^\nu-y^\nu), \quad{} \quad{} K_n = (\omega_{k_n}, k_n).
\end{align}
$iS^+, iS^-$ are the fermionic Wightman two-point functions and $i\Delta^+, i\Delta^-$ are the bosonic Wightman two-point functions. The subscript $x$ in the Dirac operator is used to emphasize that differentiation is carried out with respect to the $x$ variable only. In the massless limit $(m\rightarrow 0)$, we find that the two-point correlation function \eqref{Two_PointFunct} reduces to 
\begin{align}
W^{(2)}(x, x^\prime)
 =  -\Tr\left[\left(S^+\left(x, x^\prime\right)\right)^2\right].
\end{align}




\subsection{Twisted spinor response function for arbitrary detector's motion}
In the massless limit, we obtain the following response function for the twisted spinor 
\begin{align}\label{GeneralRFTS}
\mathcal{F}(\Omega) &= \frac{2}{L^2}  \sum_{n,m=0}^{\infty}\int_{-\infty}^{\infty} \mathrm{d}\tau \int_{-\infty}^{\infty} \mathrm{d}
\tau^\prime  \chi(\tau^\prime)\chi(\tau)e^{-i\Omega(\tau-\tau^\prime)}\exp\left(-i\frac{2\pi}{L}\left(t(\tau)-t(\tau^\prime)\right)-i\epsilon)\right) 
 \nonumber \\ &\qquad{} \times \exp \left[ -i\frac{2\pi}{L}\Big[(n+m)\left(t(\tau)-t(\tau^\prime)\right) - i\epsilon) + (n-m)\left(x^1(\tau)-x^1(\tau^\prime)\right)   \Big]     \right].
\end{align}
Clearly this expression exhibits no infrared divergence.




\subsection{Oscillator modes response function for untwisted spinor}
Consider now the untwisted spinor, whose frequencies are given by
\begin{equation}
\omega_n = \left[\left(\frac{2\pi}{L}n\right)^2 + m^2                     \right]^{1/2}. \nonumber
\end{equation}
Thus as $(m\to 0)$, $\omega_n =0$ when $n= 0$, leading to $\psi(t,x^1) \to \infty$ since the field modes are inversely proportional to $\omega_n$ as shown in \eqref{fieldsmodes}. We identify this vanishing frequency mode as the zero mode, which clearly is ill-defined. To deal with this, 
we split the fermionic fields as
\begin{subequations}
\begin{align}
\psi(t,x^1) &= \psi^{osc}(t,x^1) + \psi^{zm}(t), \\
\overline{\psi}(t,x^1) &= \overline{\psi}^{osc}(t,x^1) + \overline{\psi}^{zm}(t), 
\end{align}
\end{subequations}
where the oscillator modes, $i.e, n\not= 0$, corresponding to $\left( \psi^{osc}(t,x^1),
  \overline{\psi}^{osc}(t,x^1) \right),$
 contains the Fourier modes that are not spatially constant. And the zero mode $(n= 0)$ 
corresponding to $ \left( \psi^{zm}(t), \overline{\psi}^{zm}(t)       \right)$ contains the 
Fourier modes that are spatially constant. These modes are normalised via the anticommutator relations
\begin{subequations}\label{AntiCommu}
\begin{eqnarray}
\Big\{\psi^{osc}_a(0,x),\psi^{osc}_b(0,y) \Big\}&=  &\Big\{\psi_a^{osc^{\dagger}}(0,x),
\psi_b^{osc^{\dagger}}(0,y) \Big\} = 0, \\
\Big\{\psi^{zm}_a(0),\psi^{zm}_b(0) \Big\}&=  &\Big\{\psi_a^{zm^{\dagger}}(0),
\psi_b^{zm^{\dagger}}(0) \Big\} = 0, \\
\Big\{\psi^{osc}_a(0,x),\psi^{osc^{\dagger}}_b(0,y) \Big\} &= & \delta_{ab}\delta(x-y) - \frac{\delta_{ab}}{L}  \\
\Big\{\psi^{zm}_a(0),\psi^{zm^{\dagger}}_b(0) \Big\} &= &\frac{\delta_{ab}}{L},
\end{eqnarray}
\end{subequations}
where all the anticommutator relations between mixed field modes vanish. These relations are defined in agreement with \eqref{AntiCommRel}.

The oscillator modes contribution to the Minkowskian vacuum is
\begin{align}\label{OscMcont12}
\left\langle 0 \right|\psi^{osc}\left(x\right)\overline{\psi}^{osc}\left(y\right)\left|0 \right\rangle
&\equiv iS^+_{n\not=0}(x,y),  
\end{align}
where we define $ iS^+_{n\not =0}$ as the oscillator modes fermion Wightman function. Using this result, we then obtain the oscillator mode contribution to the detector response function for the untwisted spinor 
\begin{align}\label{GeneralforOscRFunctUS}
\mathcal{F}^{osc}(\Omega) &= \frac{2}{L^2}  \sum_{n,m=1}^{\infty}\int_{-\infty}^{\infty} \mathrm{d}\tau \int_{-\infty}^{\infty} \mathrm{d}
\tau^\prime  \chi(\tau^\prime)\chi(\tau)e^{-i\Omega(\tau-\tau^\prime)} \exp\Bigg[ -i\frac{2\pi}{L}\Big[(n+m)\big(t(\tau)\nonumber \\ &\qquad{}-t(\tau^\prime)\big)   - i\epsilon) + (n-m)\left(x^1(\tau)-x^1(\tau^\prime)\right) \Big]\Bigg],
\end{align}
which clearly exhibits no infrared divergence.




\subsection{Zero mode contribution to the response function for the untwisted spinor}
The zero mode two-point correlation is spatially independent, and moreover this mode
 does not begin in a Fock vacuum. It reads
\begin{align}\label{eq:erl7}
W^{(2)}\left(x(\tau),x(\tau^{\prime})\right)&= \left<State \right|\overline{\psi}^{zm}(t\left(\tau)\right)\psi^{zm}(t\left(\tau)\right)
\overline{\psi}^{zm}(t\left(\tau^{\prime})\right)\psi^{zm}(t\left(\tau^{\prime})\right)\left|State\right>.
\end{align}
The zero mode Hamiltonian is such that $[ H^{zm},\psi^{zm}] = 0$ $\to \psi^{zm} = const.$, therefore \eqref{eq:erl7} reduces to
\begin{equation}
\left<State \right|\overline{\psi}^{zm}\psi^{zm}\overline{\psi}^{zm}\psi^{zm}\left|State\right> = \left<State \right|
\left(\overline{\psi}^{zm}\psi^{zm}\right)^2\left|State\right>.
\end{equation}
We can write $\psi^{zm}$ as a state with two Fermi degrees of freedom as 
in Ref.~\refcite{Mark-Claudio}, 
\begin{align}
\psi^{zm} = \frac{1}{\sqrt L}
\begin{pmatrix}Q_1\\Q_2
\end{pmatrix},
\end{align}
where the Fermi degrees of freedom $Q_1$ and $Q_2$ have the following properties 
\begin{subequations}\label{FermiDeg}
\begin{align}
Q_1\left|0 \right> = Q_2\left|0 \right> &= 0 ;  \ \ \ \ \ \ \ \ \ \    Q_1^{\dagger}\left<0 \right| = Q_2^{\dagger}\left<0 \right| = 0,   \\
Q_1Q_1= Q_2Q_2 &= 0 ;  \ \ \ \ \ \ \ \ \ \    Q_1^{\dagger}Q_1^{\dagger} = Q_2^{\dagger}Q_2^{\dagger} = 0,  \\
Q_1Q_1^{\dagger} +Q_1^{\dagger}Q_1 &= 1 ;  \ \ \ \ \ \ \ \ \ \    Q_2Q_2^{\dagger} +Q_2^{\dagger}Q_2 = 1.
 \end{align}
\end{subequations}
The normalized Heisenberg picture zero-mode quantum state can now be constructed as a linear combination of the four independent states
\begin{equation}\label{ZeroModeState}
\left|State\right> = \frac{1}{|a|^2+|b|^2 +|c|^2+|d|^2}\left(a\left|0\right>   +   bQ_1^{\dagger}\left|0\right>   +    cQ_2^{\dagger}\left|0\right>   +        dQ_1^{\dagger}Q_2^{\dagger}\left|0\right>\right),
\end{equation}
where $a, b, c, d$ are complex numbers. We then obtain the zero mode two-point correlation function
\begin{align}
 W_{zm}^{(2)}=\left<State \right|\left(\overline{\psi}^{zm}\psi^{zm}\right)^2\left|State\right> = \frac{1}{L^2}\frac{|b|^2 +|c|^2}{|a|^2+|b|^2 +|c|^2+|d|^2}\equiv \frac{\theta}{L^2}, 
\end{align}
where $ 0 \leq \theta \leq 1$. Thus, the zero mode two-point correlation function is well defined. Consequently, its contribution to the detector response function reads
\begin{align}\label{ZMRF}
\mathcal{F}^{zm}(\Omega) &= \frac{\theta}{L^2} \int_{-\infty}^{\infty} \mathrm{d}\tau \int_{-\infty}^{\infty} \mathrm{d} \tau^\prime  \chi(\tau^\prime)\chi(\tau)e^{-i\Omega(\tau-\tau^\prime)} =\frac{\theta}{L^2}|\hat{\chi}(\Omega)|^2,
\end{align}
where $\hat{\chi}(\Omega)$ is the Fourier transform of $\chi(\tau)$.  
Moreover we find that the two-point correlation function for the untwisted spinor fully decouples to
\begin{equation}
W^{(2)}\left(x\left(\tau\right),x\left(\tau^{\prime}\right)\right) = W_{osc}^{(2)}\left(x\left(\tau\right),x\left(\tau^{\prime}\right)\right) + W_{zm}^{(2)}.
\end{equation}
Consequently the response function for the untwisted spinor is given by
\begin{equation}\label{eq:erl3}
\mathcal{F}(\Omega) = \mathcal{F}^{osc}(\Omega) + \mathcal{F}^{zm}(\Omega), 
\end{equation}
where the zero mode component is independent of the detector's trajectory.

\section{Conclusion}

In this paper, we examined a UDW particle detector that is coupled linearly to the scalar
density of a massless Dirac field on the static cylindrical quotient of the (1+1)-dimensional Minkowski spacetime. We showed that the twisted spin structure exhibits no infrared divergence and we
obtained its response function for arbitrary detector's motion on the cylinder. We then quantized
the zero mode present for the untwisted spin structure by treating it as a state with two Fermi
degrees of freedom. We showed that the zero mode contributes to the untwisted spin structure response
function by a state-dependent but well defined and controllable amount.

Further detail on these results, including applications to specific
trajectories and the recovery of Minkowski space results in the limit
of large $L$, will be given in a forthcoming paper.


\begin{thebibliography}{00}


\bibitem{Unruh:1976db} 
  W.~G.~Unruh,
 {\em Phys.\ Rev.} {\bf D14}, 870 (1976).


\bibitem{Dewitt}
B.~S. DeWitt,
``Quantum gravity: the new synthesis'', 
in {\it General Relativity: an Einstein centenary survey}, 
edited by S.~W.~Hawking and W.~Israel (Cambridge University Press, Cambridge, 1979)
680. 

\bibitem{Takagi:1986kn} 
  S.~Takagi,
  ``Vacuum noise and stress induced by uniform acceleration: Hawking-Unruh effect in Rindler manifold of arbitrary dimension,''
{\em  Prog.\ Theor.\ Phys.\ Suppl.}  {\bf 88}, 1 (1986).


\bibitem{birrell-davies}
N.~D. Birrell
and
P.~C.~W. Davies, 
{\it Quantum Fields in Curved Space\/}  
(Cambridge University Press, 1982).


\bibitem{Louko:2007mu} 
  J.~Louko and A.~Satz,
  ``Transition rate of the Unruh-DeWitt detector in curved spacetime,''
  {\em Class.\ Quant.\ Grav.}  {\bf 25}, 055012 (2008)
  [arXiv:0710.5671 [gr-qc]].



\bibitem{louko-satz-spatial}
J.~Louko and A.~Satz,
``How often does the Unruh-DeWitt detector click? 
Regularisation by a spatial profile,''
{\em Class.\ Quant.\ Grav.} {\bf 23}, 6321 (2006)
[arXiv:gr-qc/0606067].


\bibitem{Barbado:2012fy} 
L.~C.~Barbado and M.~Visser,
``Unruh-DeWitt detector event rate for trajectories with time-dependent acceleration,''
{\em Phys.\ Rev.}  {\bf D86}, 084011 (2012)
[arXiv:1207.5525 [gr-qc]].


\bibitem{Juarez-Aubry:2014jba} 
  B.~A.~Juárez-Aubry and J.~Louko,
  {\em Class.\ Quant.\ Grav.} {\bf 31}, 245007 (2014)


\bibitem{Martin-Martinez:2014qda} 
E.~Mart\'in-Mart\'inez and J.~Louko,
``Particle detectors and the zero mode of a quantum field,''
{\em Phys.\ Rev.}  {\bf D90}, 024015 (2014)
[arXiv:1404.5621 [quant-ph]].


\bibitem{Mark-Claudio}
Marc Henneaux
and
Claudio Teitelboim, 
{\it Quantization of Gauge Systems\/}  
(Princeton University Press, 1992).


\end{thebibliography}
\end{document}